\documentclass{aa}
\usepackage{graphics}
\usepackage{graphicx}
\usepackage{amssymb}                                                             
\usepackage{latexsym}

\begin{document}
\title{The nature of super metal rich stars \thanks{Based on observations 
obtained at the McDonald Observatory.}
}
 
\subtitle{Detailed abundance analysis of 8 super-metal-rich star
candidates}
 
\author{Sofia Feltzing \inst{1} 
Guillermo Gonzalez \inst{2}
  }

\offprints{Sofia Feltzing  sofia@astro.lu.se}
 
\institute{
Lund Observatory, Box 43, SE-221 00 Lund, Sweden
\and Astronomy Department, University of Washington, P. O. Box 
351580, Seattle, WA 98195 USA
            }

\date{}

\abstract{
We provide detailed abundance analyses of 8 candidate super-metal-rich 
stars. Five of them are confirmed to have 
[Fe/H]$>$0.2 dex, the generally-accepted limit for super-metal-richness. 
Furthermore, we derive abundances of several elements and find that the
stars 
follow trends seen in previous studies of metal-rich stars.
Ages are estimated from isochrones and velocities calculated.
We find that there do exist very metal-rich stars that are older than 10 
Gyr. This is contrary to what is found in several recent studies of the 
galactic age-metallicity relation. This is tentative evidence that there 
might not exist a one-to-one relation between age and metallicity for
all stars. This is not surprising considering the current models of the 
independent evolution of the different galactic components. 
We also find that one star, HD 182572,  could 
with $\sim$ 75\% chance be a thick disk star with, for the thick disk,
an extremely high metallicity at 0.34 dex. This star is, intriguingly, 
also somewhat enhanced in the $\alpha$-elements.
      }
\maketitle

\section{Introduction}

The very metal-rich dwarf stars in the solar neighbourhood have
historically not attracted as much attention as the more metal-poor
(solar  like and halo stars) stars which tell us about the early
phases of the chemical evolution of our galaxy. The properties of
metal-rich stars are important when we try to interpret integrated
spectra from metal-rich stellar populations, such as the Bulge and
giant elliptical galaxies. A small group of so called super-metal-rich 
(SMR) stars have played a significant role in shaping the conceptions of
such populations. Famous examples are the dwarf HD32147 (HR1614) and
the giant  $\mu$ Leonis. In the  review by Taylor (1996) - the latest
paper in a long series started in the  1960s - SMR stars are discussed
in great detail, in particular,  the reality of extremely high
[Fe/H]. Taylor found that no giant  star fulfills the criteria for
SMR-ness that he sets and only a handful of  dwarf stars do, and that most
of them are candidates rather than  firm members in this class. $\mu$
Leonis has, however, been studied by several groups using
high-resolution spectroscopy, a recent example being Smith \& Ruck
(2000), who find that the star is indeed  super-metal-rich with
[Fe/H]$ = +0.29$. Thus, the question of the reality of
super-metal-rich giants is still very much 
alive  and each case has to be judged on its own.

The exact definition of super-metallicity has, as reviewed by Taylor (1996),
varied. Spinrad \& Taylor (1969) adopted $+0.2$ dex  as
the lower limit, based on the overall metallicity of the Hyades,
which they found to be +0.2 dex. The metallicity for the Hyades has
recently been revised (Taylor 1994 and Cayrel de Strobel 1997)  to
+0.1 dex. Even values as low as 0.0 dex have been quoted. This has
resulted in classes of stars that sometimes are regarded as SMR and sometimes
not. Taylor rectified this unsatisfactory  situation by
adopting the original $+0.2$ dex as the threshold on the  grounds that
no giant stars had been shown to have a metallicity  higher than this
value (but see Castro et al. 1997 and Smith \& Ruck  2000). Taylor (1996)
 defines  a star to be SMR if it has  [Fe/H]$ >$ 0.20 with 95\% confidence.
He
also adopts  [Fe/H], i.e. the iron abundance, as the measure of
``metallicity'' rather  than the more general [Me/H]. As an aside one
may note that a second  terminology is also in use - Very Strong-Lined
(VSL) star. This term implies  just that the star has strong lines and
might therefore be a SMR candidate.  This is a particularly useful
term when working with low resolution spectra.

SMR stars have attracted more attention recently due to their possible
connection with extra-solar planets, e.g. Gonzalez (2000 and
references cited  therein), Fuhrmann et al. (1997, 1998). Gonzalez
(2000) has shown that the solar-type parent stars of extra-solar
planets are more metal-rich on  average compared to the general field star
population. In particular, the  very short period systems are either
above the SMR  limit or near it.  By  comparing them to the SMR
stars we may gain insight as to the relationship  between planets in
short-period orbits and the SMR-ness of the parent star. 

A few other recent studies have targeted known SMR candidates
 and stars with high [Me/H] (as derived from
photometry): Feltzing \& Gustafsson (1998), Castro et al. (1997), and
McWilliam \& Rich  (1994). In general the abundance
ratios  seem to continue the trends of the disk population. However,
no detailed  theoretical predictions for Galactic chemical evolution
exists for [Fe/H]  $>$ 0.2 dex, so the interpretation of the observed
abundance trends for  metal-rich stars is still pending.

The combination of abundance ratios with kinematical data may give us
additional clues. For example, we can study stars on highly eccentric
orbits  which trace the evolution in the Galactic disk closer to the
Galactic centre.  Not much is known about these stars, but there are
some very intriguing  observations: Barbuy \& Grenon (1990) found that
dwarf stars on very  eccentric orbits contained much more oxygen than
what was expected from  standard models of Galactic chemical evolution
of the disk, and Edvardsson  et al. (1993) found large spreads and
``upturns'' for certain elements, Na, Si,  Ti, Al, for stars with $0.0
{\rm dex} < {\rm [Fe/H]}< 0.2$ dex. The trends  for Na, Si and Ti were
confirmed up to $\sim 0.4$ dex by Feltzing \&  Gustafsson (1998). They
concluded that the ``upturn'' in Na abundances relative to Fe  is not
due to a mixture of stars born at different distances from the
Galactic centre.

In this paper we investigate, by means of detailed spectroscopic 
analyses, the metallicities as well as the abundance of several elements 
for 8 dwarf stars selected from the meticulous review of 
 SMR candidates by Taylor (1996).

The paper is organized as follows: in Sect. 2 and 3 we detail the
observations and the selection of program stars, as well as
reductions and measurements; Sect. 4 discusses the detailed abundance
analysis, Sect. 5 presents the abundances element by element, in Sect. 6
we derive ages for the stars and discuss the age-metallicity relation in
the solar neighbourhood, Sect. 7 discusses the kinematics of the stars
in our sample and which galactic component they belong to, Sect. 8
provides
a short discussion of the SMR-planet connection and,  finally, Sect. 9
summarizes our findings.

\section{Selection of program stars and observations}

\begin{table*}
\caption[]{Stellar parameters for the program stars. Magnitudes,
colours and parallaxes are from the Hipparcos catalogue (ESA 1997). The 
spectral types are from the SIMBAD database. Effective temperatures, 
surface gravities, [Fe/H] and microturbulence parameters as 
determined in this study.
}
\begin{tabular}{lllccccccccccccccc}
\hline\noalign{\smallskip}
 & & Sp.T. & $V$ & $B-V$ & $\pi$ & $\sigma_{\pi}
(\sigma_{\pi}/\pi)$&$T_{\rm eff}$ 
& log{\it g} & {[Fe/H]} & $\xi_t$\\
 & & & & & mas &&  & & & km~s$^{\rm -1}$ & & & &\\
\noalign{\smallskip}
\hline\noalign{\smallskip}
HD10780  &  HR511 & K0V     & 5.63 & 0.804 & 100.24 & 0.68(0.007)& 5300 &
4.13 & -0.02 & 1.00 \\
HD32147  & HR1614 & K3V     & 6.22 & 1.049 & 113.46 & 0.82(0.007)& 4680 &
4.00 & 0.28  & 0.50 \\
HD99491  &HR4414A & K0IV    & 6.49 & 0.778 &  56.59 & 1.40(0.025)& 5300 &
4.12 & 0.20 & 1.00 \\
HD104304 & HR4587 & G9IV    & 5.54 & 0.760 &  77.48 & 0.80(0.01)& 5400 &
4.12 & 0.16 & 1.15 & \\
HD121370 & HR5235 & G0IV    & 2.68 & 0.580 &  88.71 & 0.75(0.008)& 6000 &
3.66 & 0.25 & 2.00\\
HD145675 &        & K0V     & 6.61 & 0.877 &  55.11 & 0.59(0.011)& 5300 &
4.50 & 0.47 & 1.00& \\
HD182572 & HR7373 & G8IV    & 5.17 & 0.761 &  66.01 & 0.77(0.012)& 5400 &
4.00 & 0.35 & 1.10 \\
HD196755 & HR7896 & G5IV+   & 5.07 & 0.702 &  33.27 & 0.82(0.02)& 5700 &
4.00 & 0.02 & 1.50 & \\
\noalign{\smallskip}
\hline\noalign{\smallskip}
$\alpha$ Cen A\footnotemark[1] &&  & & & & &5830 & 4.34 &0.24 & 1.09\\
$\alpha$ Cen B\footnotemark[1] &&  & & & & &5225 & 4.51 &0.23 & 1.00\\ 
\noalign{\smallskip}
\hline\noalign{\smallskip}
 \multicolumn{10}{l}{1. $T_{\rm eff}$, $\log g$ and $\xi_t$  
from Neuforge-Verheeke \& Magain (1997)}\\
\label{parameters}
\end{tabular}
\end{table*}

\stepcounter{footnote}

Stars observable from the Northern Hemisphere were selected from
Taylor's 1996  list of candidate SMR dwarf and subgiant  stars. Spectra
were obtained with  the Sandiford cassegrain echelle spectrograph
(McCarthy et al. 1993) attached  to the 2.1 m Struve telescope at
McDonald Observatory during three runs: 1996  June, 1996 August and
1996 December. Exposure times were typically 5 minutes  each,
resulting in signal-to-noise (S/N) ratios per pixel averaging near
250.  A spectrum of a Th-Ar lamp was obtained following each star
spectrum,  ensuring accurate wavelength calibration. The resolving
power (measured on the  Th-Ar emission line spectra) averages near $R
= 45,000$.

\section{Reductions and measurements}

\begin{figure}
\resizebox{\hsize}{!}{\includegraphics{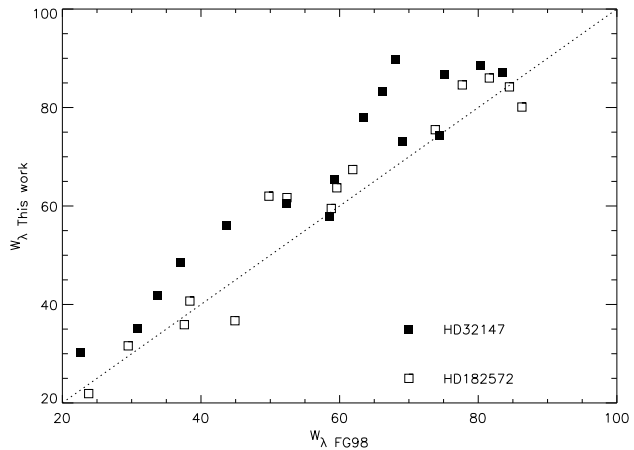}}
\caption[]{Comparison of measurements of $W_{\lambda}$ ({\AA}) 
in this work and 
in Feltzing \& Gustafsson (1998) for HD 32147 and HD 182572 }
\label{jmfr_wl}
\end{figure}

The spectra were reduced with the standard software available within the 
{\sc ccdred} and {\sc echelle} packages of {\sc noao iraf}\footnote{IRAF
is
 distributed by National Optical Astronomy Observatories, operated by
 the Association of Universities for Research in Astronomy, Inc., under 
contract with the National Science Foundation, USA.}. The steps included 
bias subtraction, flat fielding, extraction of one-dimensional spectra, 
wavelength calibration, and continuum normalization.  Additional details 
concerning the quality of the data resulting from the Sandiford
spectrograph 
can be found in Gonzalez \& Lambert (1996) and Gonzalez (1998).

Equivalent widths ($W_{\lambda}$) were measured using the {\sc splot} 
task in {\sc iraf}. The lines were measured both by simply integrating
the line and also by fitting a Gaussian to the line profile. Most lines
were measured twice and some up to four times due to overlap of the 
spectral orders. As the final adopted value of $W_{\lambda}$ we used 
the mean of the measurements. In these cases the measurement errors are
typically
no more than a few percent. 

In Fig. \ref{jmfr_wl} we compare the measured values of $W_{\lambda}$ 
for HD 32147 and HD 182572 with those measured by Feltzing \& 
Gustafsson (1998). For HD 182572 the agreement 
is good, while for HD 32147, our coolest star, we measure significantly
larger $W_{\lambda}$. This difference is most likely due to the
lower resolution used in this work. See also the two examples of
stellar spectra shown in Fig. \ref{feii_lines} from which it is clear
that HD 32147, but also to some extent HD 145675, shows a much 
richer spectrum than the other stars. Since these stars are cool, there
will naturally be more molecular lines and low-excitation atomic lines 
that will cause blending problems.

\section{Analysis}

We have performed a standard Local Thermodynamic Equilibrium (LTE)
analysis, strictly differential with respect to the Sun, to derive
chemical
abundances from the measured values of $W_{\lambda}$. Spectrum 
synthesis was not employed in the present study.

\subsection{Model atmospheres}

To generate the model atmospheres we used the MARCS program, first
described by Gustafsson et al. (1975). The program has been further
developed and updated in order to handle the line blanketing of millions 
of absorption lines more accurately (Asplund et al. 1997). The following 
assumptions enter into the calculation of the models: the atmosphere is 
assumed to be plane-parallel and in hydrostatic equilibrium, the total
flux 
(including mixing-length convection) is constant, the source function is 
described by the Planck function at the local temperature with a
scattering 
term, the populations of different excitation levels and ionization stages
are 
governed by LTE.

Since our analysis is strictly differential relative to the Sun, we have
used a 
solar model atmosphere calculated with the same program as the stellar 
models -- this in order to keep the analysis truly differential and 
thus in spite of the fact that the empirically derived 
Holweger-M\"uller model better reproduces the solar observed limb
darkening 
(Blackwell et al. 1995).

\subsection{Line data}

\begin{figure}
\resizebox{\hsize}{!}{\includegraphics{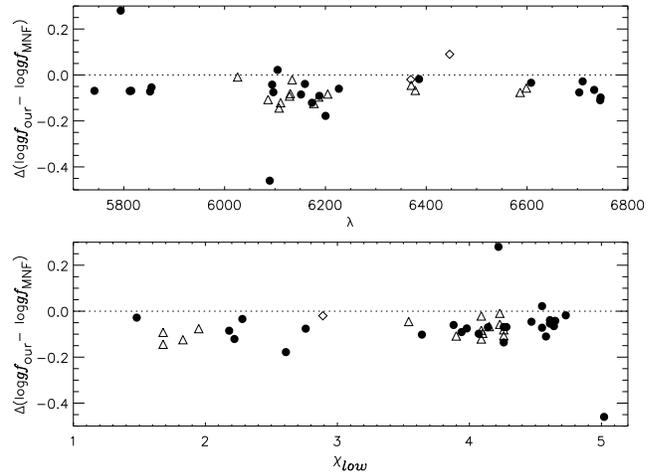}}
\caption[]{Comparison of $\log gf$-values as derived in this study and in
Neuforge-Verheecke \& Magain (1997) study of $\alpha$ Cen A and B.
Fe\,{\sc i} lines are denoted by
$\bullet$, 
Fe\,{\sc ii} by $\Diamond$ and Ni\,{\sc i} lines by $\bigtriangleup$.}
\label{comp_loggf}
\end{figure}

Since we did not have observations of the solar spectrum for all of
the lines available in the stellar spectra, we measured solar line
strengths in 
the Kurucz et al. (1984) Solar Flux Atlas. The spectrum from the
Flux Atlas was first degraded by binnning and then convolved with a 
Gaussian profile to match the instrumental profile. To decide on the exact
values of the convolution, we used three portions of a spectrum of
reflected sunlight from Vesta. The Flux Atlas spectrum was convolved
and then compared with the Vesta spectrum. The goodness of the fit was
decided upon by inspection. The values of $W_{\lambda}$ for all our lines 
were measured in the degraded spectrum. They were then used to determine 
the astrophysical $\log gf$ values, Table \ref{linelist}.

We consider different line broadening mechanisms in our
calculations; van der Waals damping, radiation damping, thermal
Doppler broadening and  broadening by microturbulence. The van der
Waals broadening is calculated with the classical Uns\"old
formula. Enhancement factors to this value were compiled from the
literature and are given in Table \ref{linelist}.  For Fe we use
values from Hannaford et al. (1992) and Holweger et al. (1991), for
Ca,  and for V from Neuforge
(1992). For the remaining lines we use a value of 2.5, according to
M\"ackle et al. (1975). The values used for the enhancement factor do
not, in general, influence the results, e.g. a change from 2.5 to 1.4 does
not alter most abundances by more than 0.01 dex. 

We have also compared our $\log gf$-values derived from the solar spectrum 
with those derived in a similar way, but using a Holweger-M\"uller solar 
model, by Neuforge-Verheecke and Magain (1997), Fig. \ref{comp_loggf}.
Our $\log gf$-values are 0.07 dex lower for Fe\,{\sc i} lines and
0.04 dex lower for the Ni\,{\sc i} lines than those derived 
 by Neuforge-Verheecke
and Magain (1997). Considering the different approaches to the
derivation of the astrophysical $\log gf$-values we consider the
agreement good. It is also reassuring that no trends are found
either with wavelength or excitation potential, see Fig. \ref{comp_loggf}.

\subsubsection{Selection of lines}

Selecting stellar lines which are free from blends is crucial for deriving
accurate elemental  abundances. To account for telluric lines we
simply over-plotted each stellar spectrum with a spectrum of a hot star
observed during the same night as the stellar spectrum was taken and
discarded lines that were contaminated. To avoid blends from unidentified
photospheric lines the solar spectrum was carefully inspected and the
line-list by Moore consulted.

\begin{figure}
\resizebox{\hsize}{!}{\includegraphics{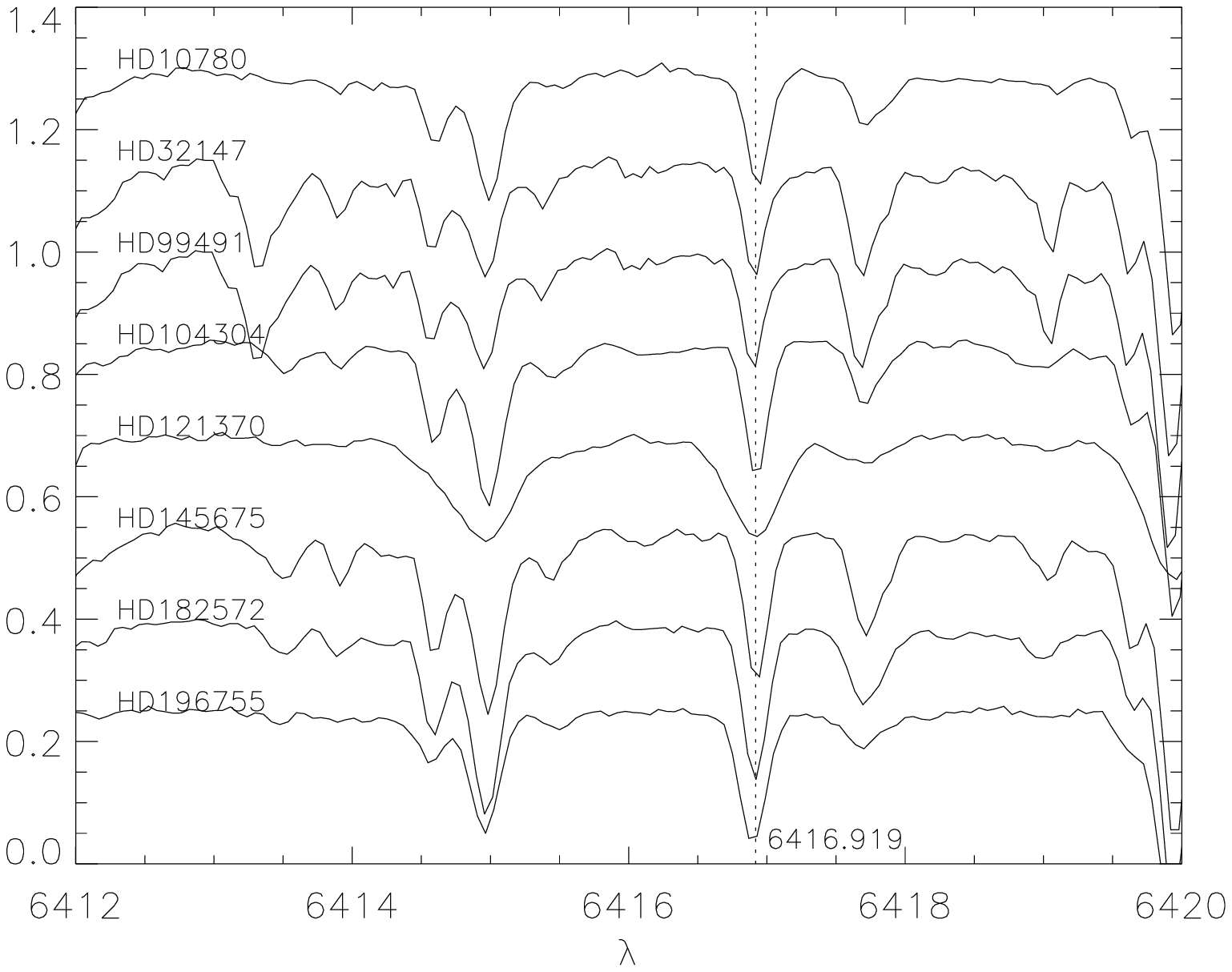}}
\resizebox{\hsize}{!}{\includegraphics{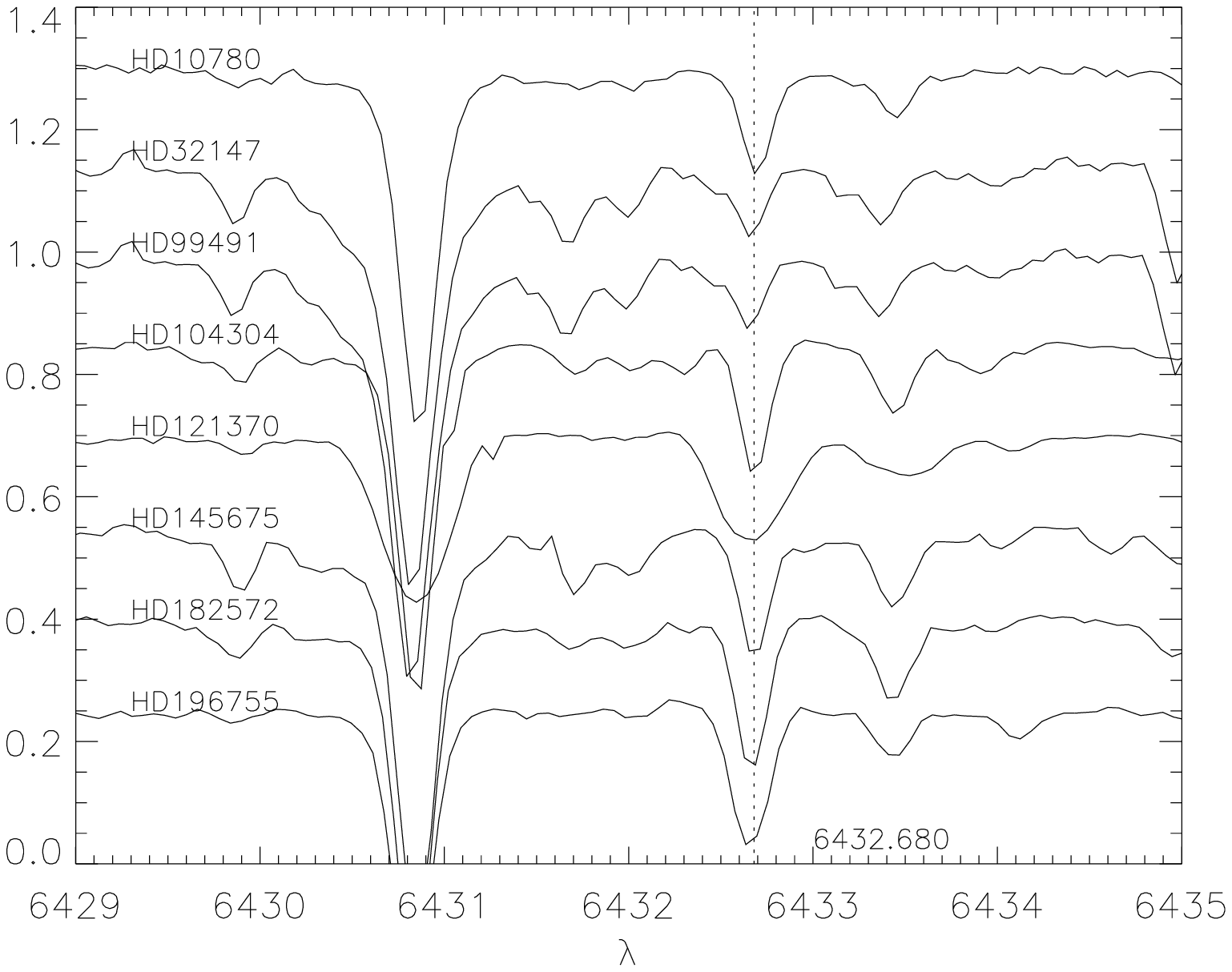}}
\caption[]{Two portions of the stellar spectra, showing
the regions around  the two 
Fe\,{\sc ii} lines at 6416.91 and 6432.68 {\AA}. The spectra have been 
arbitrarily displaced in intensity and also along the x-axis to the 
laboratory wavelengths. The positions of the Fe\,{\sc ii} lines are marked 
with dotted lines. Note the different scales on the x-axes.}
\label{feii_lines}
\end{figure}

Care in the selection of lines is also of importance for the
determination of surface gravities by means of ionization equilibrium
(i.e., abundances derived from Fe\,{\sc i} and Fe\,{\sc ii} lines give
the same iron abundance). We have inspected the shape of the Fe\,{\sc
ii} lines in all the stars when a line is observed in more than two
stars. This inspection led us to discard the lines at 6386.72 and
7449.33 {\AA}, Fig. \ref{feii_lines}. 
A line at 5823.15 {\AA} was also discarded, although
only measured in two stars, since it gave anomalously high iron
abundances and clearly suffered from blends. The line at 6416.91 {\AA}
gave rather high iron abundances in HD 10780, HD 32147 and
HD 145675. From our spectra we could not, however, conclude that this
line is compromised by a blend (see Fig. \ref{feii_lines}) and it was
therefore kept in the analysis, but only in those stars where it did
not diverge significantly.  Our final selection of lines, as well as
the parameters used in the  abundance analysis, are given in Table
\ref{linelist}. 

\subsection{Fundamental parameters of model atmospheres} 

In order to construct the stellar model atmospheres we need the
effective temperature, surface gravity, metallicity and
microturbulence for each star.  These were all derived from the
stellar spectra themselves.

\paragraph{Effective temperature}

\begin{figure*}
\resizebox{\hsize}{!}{\includegraphics{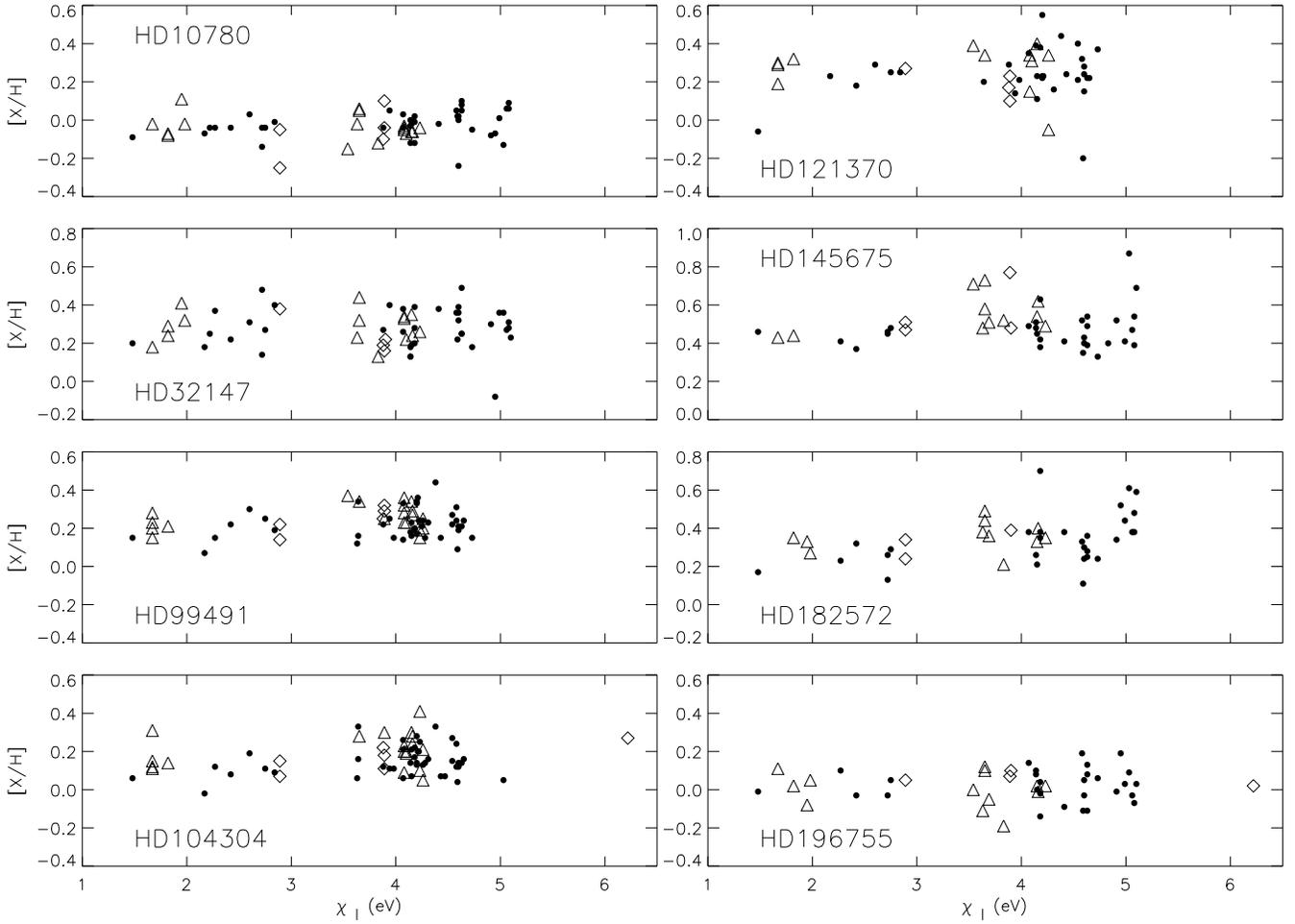}}
\caption[]{Excitation equilibrium. Fe\,{\sc i} lines are denoted by
$\bullet$, 
Fe\,{\sc ii} by $\Diamond$ and Ni\,{\sc i} lines by $\bigtriangleup$.
[X/H] 
denote the abundance relative to solar as derived from Fe\,{\sc i}, 
Fe\,{\sc ii} and Ni\,{\sc i} lines, respectively. Note the 
different ranges on the y-axes.}
\label{fe_vs_chi}
\end{figure*}

Initial estimates of effective temperatures for the stars were determined 
using the photometric calibrations by Alonso et al. (1996). These
estimates 
were iterated until excitation energy equilibrium was achieved. The 
plots from which the final temperatures were derived are shown in 
Fig. \ref{fe_vs_chi}, and the final adopted temperatures are given in 
Table \ref{parameters}. 

\paragraph{Surface gravity}

Surface gravities were determined by requiring ionization equilibrium
for Fe abundances derived from Fe\,{\sc i} and Fe\,{\sc ii}
lines. We adjusted $\log g$ until the
iron abundance derived from Fe\,{\sc i} and Fe\,{\sc ii} lines
gave the same Fe abundance, compare Fig. \ref{fe_vs_chi}.

\paragraph{Metallicity}

Our first estimates were taken from Taylor (1996). The [Fe/H] were 
iterated until the metallicity used in constructing
the atmosphere and the derived [Fe/H] agreed. 

\paragraph{Microturbulence}

The microturbulence parameter, $\xi_t$, which is introduced to account
for
unknown line broadening mechanisms, affects strong and weak absorption
lines differently.  For weak lines only the shape, and not the
$W_{\lambda}$ is affected, but for strong lines the line strength
increases when $\xi_t$ is increased. We use these trends to constrain
the value of $\xi_t$.

We started with $\xi_t = 1.00$ km/s and, after inspecting plots of [Fe/H] 
and [Ni/H] as a functions of  $W_{\lambda}/\lambda$ (reduced 
equivalent width), varied the value of $\xi_t$ until all lines, weak 
and strong, yielded the same abundance. The final values used in the 
abundance analysis are given in Table \ref{parameters}.

\subsection{Comparison/verification with calibrations of 
$uvby-\beta$ photometry}

As a further check of our final stellar parameters we have derived
$T_{\rm eff}$, $\log g$, and [Fe/H] from the self-consistent
calibration of $T_{eff}$, $\log g$ and [Fe/H] by Olsen (1984),
 Table \ref{comp_para}. The agreement is in general good. 

\begin{table}
\caption[]{Str\"omgren photometry and stellar parameters derived from the 
photometry thorough the calibration by Olsen (1984). References;
O93=Olsen (1993), O94a=Olsen (1994a), O94b=Olsen (1994b), GO= 
Gronbech \& Olsen(1997)} 
\begin{tabular}{lrrrlrrrrrrrc}
\hline\noalign{\smallskip}
 ID & b-y &$m_1$ & $c_1$& ref. &\multicolumn{3}{c}{Olsen (1984)}  \\
    &     &      &      &      & $T_{\rm eff}$& $\log g$\\
\noalign{\smallskip}
\hline\noalign{\smallskip}
HD 10780   & 0.468 & 0.316 & 0.327 & O93 & 5431 & 4.27 \\
HD 32147   & 0.601 & 0.634 & 0.236 & O94a& 4614 & 4.57 \\
HD 99491   & 0.484 & 0.335 & 0.362 & O93 & 5347 & 4.12 \\
HD 104304  & 0.469 & 0.313 & 0.345 & O94a& 5437 & 4.19 \\
HD 121370  & 0.370 & 0.202 & 0.533 & GO  & 6205 & 3.92 \\
HD 145675  & 0.537 & 0.336 & 0.438 & O93 & 4852 & 4.55 \\
HD 182572  & 0.465 & 0.299 & 0.381 & O93 & 5495 & 4.07 \\
HD 196755  & 0.432 & 0.220 & 0.381 & O94b& 5642 & 3.98 \\
\noalign{\smallskip}
\hline
\label{comp_para}
\end{tabular}
\end{table}

\subsection{Comparison with other studies}
\label{sec:comp}

\begin{table}
\caption[]{Comparison of results from this study
with those of Feltzing \& Gustafsson (1998), FG98, for HD 32147 and HD 182572} 
\begin{tabular}{lrrrrrrrrrrc}
\hline\noalign{\smallskip}
    &\multicolumn{2}{c}{HD 32147} & \multicolumn{2}{c}{HD 182572}\\
 & This work  & FG98 & This work & FG98 \\
\noalign{\smallskip}
\hline\noalign{\smallskip}
Al\,{\sc i}  & 0.48 & 0.25  & 0.55  & 0.53 \\  
Si\,{\sc i}  & 0.36 & 0.48  & 0.49  & 0.51 \\
Ca\,{\sc i}  &      & 0.01  &   & 0.42 \\
Sc\,{\sc ii} & --   & 0.49  & 0.36  & 0.36 \\
Ti\,{\sc i}  & 0.66 & 0.11  & 0.32  & 0.50 \\
V\,{\sc i}   & 0.95 & -0.18 &   &  \\
Cr\,{\sc i}  & 0.50 & 0.10  & 0.40  & 0.43 \\
Fe\,{\sc i}  & 0.28 & 0.22  & 0.34  & 0.42 \\
Fe\,{\sc ii} & 0.24 & 0.61  & 0.32   & 0.08 \\
Co\,{\sc i} & 0.56 & 0.39  & 0.47  & 0.58 \\
Ni\,{\sc i}  & 0.29 & 0.57  & 0.36  & 0.46 \\ 
\noalign{\smallskip}
\hline
\label{comp}
\end{tabular}
\end{table}

\begin{table}
\caption[]{Comparison for HD 121370
of results from this study
with those of Edvardsson et al. (1993). The second line for 
Fe\,{\sc ii} give the iron abundance derived if the line 
at  6416.91 {\AA} is excluded. }
\begin{tabular}{lrrrrrrrrrrc}
\hline\noalign{\smallskip}
 & This work & Edv.93 \\
\noalign{\smallskip}
\hline\noalign{\smallskip}
Na\,{\sc i}  & 0.50 & 0.45\\
Si\,{\sc i}  & 0.40 & 0.31  \\
Ca\,{\sc i}  &      & 0.23  \\
Ti\,{\sc i}  & 0.22 & 0.32  \\
Fe\,{\sc i}  & 0.24 & 0.19  \\
Fe\,{\sc ii} & 0.19 & 0.25  \\
             & 0.22 &       \\
Ni\,{\sc i}  & 0.31 & 0.30  \\
\noalign{\smallskip}
\hline
\label{comp_5235}
\end{tabular}
\end{table}

The stars in our study have been included in few, if any,  abundance
studies. However, HD 32147 and HD 182572 have been  extensively
studied.  HD 32147 has been especially difficult to analyze,  because
it is a cool K dwarf star with strong lines. This is amply exemplified
by the comparison of our $W_{\lambda}$ measurements and abundances
with those of Feltzing \& Gustafsson (1998). In Table \ref{comp} we
compare  our results to theirs. As expected (from the comparison of
$W_{\lambda}$)  the abundances for HD 32147 are larger in our study
than in theirs. In this  work we impose ionization equilibrium in
order to derive the surface  gravity of the star. This affects in
particular the abundances derived from HD  32147, but also some of the
species, i.e. Fe, Co and Ni, for HD 182572.  For a discussion of stellar 
abundances in K dwarf stars, that for HD 32147 supersedes the current 
analysis, we refer the reader to Thor\'en \& Feltzing (2000).

Gonzalez et al. (1999) found HD 145675 (14 Her) to have [Fe/H] of
$0.50\pm0.05$, using a spectrum with nearly twice the
resolution as ours. Nevertheless, it is in good agreement with our
$0.47$ dex estimate with a line-to-line scatter of 0.11 dex derived using
30 lines.

For HD 104304 Fran\c cois (1988) found [Fe/H] $= 0.16$ and [S/H] $=
0.59$; our estimate of [Fe/H] $= 0.17$ is in excellent agreement.  We 
derive an [S/H] value lower by 0.10 dex, but since our
result is based on  one fairly weak S\,{\sc i} line, we consider this
to be a good agreement.

Morell et al. (1992) derived Fe and Th abundances for a group
of  stars in order to estimate their ages. For HD 182572 and HD 196755
they  derived [Fe/H] $= +0.3$ and $+0.1$ dex, respectively. This is in
reasonable  agreement with our results.

Edvardsson et al. (1993) analyzed 189 dwarf and subgiant stars with
[Fe/H] up to $+0.25$ dex, including HD 121370. The agreement  between
the two studies is very good,  Table
\ref{comp_5235}. Also, the stellar parameters agree well. These
different  comparisons give us confidence that our analysis is
satisfactory.

\subsubsection{$\alpha$ Cen A and B}
\label{sect:alphacenab}

As a final test of our analysis method and its compatibility with the
analysis procedures adopted by other groups, we derived elemental
abundances for the stars in the nearby triple system $\alpha$ Centauri
from  the equivalent widths published by Neuforge-Verheecke \& Magain
(1997).  They observed the two stars (components A and B) with the
CAT-telescope at La  Silla with a resolution of 100,000 and a final
$S/N \sim 550$ and derived stellar abundances as well as stellar
parameters self-consistently from the spectra. Using their published
equivalent widths as well as their $\log gf$ and damping parameters
with our set of model atmospheres and programs, we derive almost the
same abundances for all elements with lines that are not affected by
hyperfine splitting, see  Table \ref{comp_cenab}. In fact for most of
those elements taking the  hyperfine structure in the line profile
into account makes very little  difference in the derived
abundances. This is true in particular for Al.

\begin{table*}
\caption[]{Comparison of abundances for $\alpha$ Cen A and B derived by 
Neuforge-Verheeke \& Magain (1997) and in this work using their equivalent
widths and
models as described in Sect. \ref{sect:alphacenab}.}
\begin{tabular}{lrrrrrrrrrrc}
\hline\noalign{\smallskip}
    & \multicolumn{3}{c}{$\alpha$Cen A} & \multicolumn{3}{c}{$\alpha$Cen
B}\\
El. & \# lines& This work & NVM97  &  \# lines& This work & NVM97 \\
    &         & [X/H]$\pm$sc. & [X/H]$\pm$ error & & [X/H]$\pm$sc. &
[X/H]$\pm$ error \\
\noalign{\smallskip}
\hline\noalign{\smallskip}
O\,{\sc i}   &  3 &  0.20  0.07 & 0.21 0.06 &      &             &\\
Al\,{\sc i}  &  1 &  0.23  0.00 & 0.24 0.04 &   1  &  0.27  0.00 & 0.24
0.05\\
Si\,{\sc i}  &  3 &  0.26  0.02 & 0.27 0.03 &   3  &  0.30  0.00 & 0.27
0.04\\
Ca\,{\sc i}  &  5 &  0.21  0.05 & 0.22 0.03 &   5  &  0.23  0.05 & 0.21
0.05\\
Sc\,{\sc ii} &  1 &  0.35  0.00 & 0.25 0.05 &   1  &  0.36  0.00 & 0.26
0.04\\
Ti\,{\sc i}  & 15 &  0.22  0.04 & 0.25 0.03 &  13  &  0.26  0.07 & 0.27
0.06\\
V\,{\sc i}   &  4 &  0.22  0.04 & 0.23 0.05 &   4  &  0.40  0.07 & 0.32
0.08\\
Cr\,{\sc i}  & 11 &  0.20  0.03 & 0.24 0.02 &  12  &  0.25  0.03 & 0.27
0.04\\
Cr\,{\sc ii} &  2 &  0.25  0.02 & 0.26 0.03 &   1  &  0.29  0.00 & 0.26
0.09\\
Fe\,{\sc i}  & 69 &  0.24  0.06 & 0.25 0.02 &  65  &  0.23  0.05 & 0.24
0.03\\
Fe\,{\sc ii} &  4 &  0.25  0.03 & 0.25 0.02 &   4  &  0.27  0.04 & 0.25
0.02\\
Co\,{\sc i}  &  3 &  0.29  0.04 & 0.28 0.04 &   3  &  0.39  0.02 & 0.26
0.04\\
Ni\,{\sc i}  & 26 &  0.29  0.05 & 0.30 0.03 &  25  &  0.31  0.06 & 0.30
0.02\\
Y\,{\sc ii}  &  1 &  0.36  0.00 & 0.20 0.05 &   1  &  0.30  0.00 & 0.14
0.05\\
Eu\,{\sc ii} &  1 &  0.17  0.00 & 0.15 0.05 &   1  &  0.16  0.00 & 0.14
0.05\\
\noalign{\smallskip}
\hline
\label{comp_cenab}
\end{tabular}
\end{table*}

For the A component we derive in general abundances 0.01 dex less than
Neuforge-Verheecke \& Magain (1997) and for the B component
0.02-0.03 dex higher abundances, see Table \ref{comp_cenab}.  Iron is
however 0.01 dex lower for the B component. We find this level of
agreement satisfactory considering that we use model atmospheres of
slightly different construction.

\section{Abundance results}

\begin{table*}
\caption[]{Derived elemental abundances for our program stars.  The
abundances are given in the format $[X/H] = \log(X/H)_{\star} -
\log(X/H)_{\odot}$, where X  denotes the element in question. The
ionization stages are
also indicated. The line-to-line scatter  is given for each star
and element when more than one line is used in the abundance analysis.
The error in the derived abundance due to line-to-line scatter is thus
=  line-to-line scatter /$\sqrt n_{lines}$, the number of lines used
are given in parentheses. Oxygen abundances are given separately for the
forbidden
line at 630nm and the triplet at 777nm. No star has observations of both.} 
{\setlength{\tabcolsep}{0.8mm}
\begin{tabular}{lllllllllllllll}
\hline\noalign{\smallskip}
&HD10780& HD32147 &HD99491 &HD104304&HD121370&HD145675&HD182572&HD196755\\
\noalign{\smallskip}
\hline\noalign{\smallskip}
C\,{\sc i}&    0.28           &--           &--           & --
&--           &--           &--          &--            \\
O\,{\sc i}$_{630}$ &   --             &--           &0.27         &0.37
&--           &--           &--          &--            \\
O\,{\sc i}$_{777}$  &   --             & --          &      --     &--
&    --       & 0.48 0.15 (3)& 0.62 0.10 (3) & 0.11 0.05 (3)\\
Na\,{\sc i}&   -0.03          &0.64         &0.34         &0.37
&0.50         &--           &--          &--            \\
Al\,{\sc i}&   -0.01 0.03 (3) &0.48 0.05 (3)&0.41         &0.25
&--           &0.54 0.02 (3)&0.55 0.04(3)&0.02 0.05 (3)  \\
Si\,{\sc i}&    0.03 0.05 (5) &0.36 0.15 (5)&0.35 0.10 (7)&0.27 0.08
(8)&0.40 0.14 (6)&0.61 0.07 (3)&0.49 0.18(6)&0.09 0.05 (4) \\
           &                  &             &             &             &
&0.52 0.19 (4)& \\
Ca\,{\sc i}&   0.13 0.10 (5) &
 -- & 0.18 0.08 (4) &   0.15 0.07 (4)& 0.11  & 0.21  & 0.28 & -0.02 0.23 (2) \\
S\,{\sc i}&    --             &--           &0.56         &0.49
&0.72         &--           &--          &0.38          \\
Sc\,{\sc i} &--           &--           &0.10          &0.11          &--
&--           &--           &--                 \\
Sc\,{\sc ii}&-0.12        &0.36         &--            &0.32 0.13 (3)
&0.11         &0.66         &0.36         &0.11              \\
Ti\,{\sc i} &0.10 0.07 (2)&0.66 0.18 (2)&0.17 0.10 (13)&0.11 0.10
(12)&0.22 0.22 (3)&0.62 0.03 (2)&0.32 0.03 (3)&0.12 0.09 (3)      \\
Cr\,{\sc i} &0.01 0.15 (3)&0.50 0.12 (3)&0.19 0.11 (6) &0.14 0.09 (5)
&0.22 0.04 (2)&0.42 0.04 (3)&0.40 0.01 (2)&-0.03 0.16 (2)    \\
Cr\,{\sc ii}&--           &--           &0.35          &0.26          &--
&--           &--           &--                \\
Fe\,{\sc i} &-0.02 0.07 (39)&0.28 0.11 (39)&0.22 0.08 (42)&0.15 0.08
(44)&0.24 0.14 (32)&0.47 0.11 (30)&0.34 0.14 (29)&0.02 0.09 (28)&     \\
Fe\,{\sc ii}&-0.11 0.10 (4) &0.24 0.10 (4) &0.24 0.07 (5) &0.17 0.08 (5)
&0.19 0.07 (4) &0.49 0.02 (3) &0.32 0.08 (3) &0.06 0.03 (4) &   \\
            &               &              &              &           &
&0.56 0.14 (4) \\
Co\,{\sc i} &-0.06 0.06 (4) &0.56 0.19 (4) &0.26 0.08 (5) &0.04 0.54 (6)
&0.32          &0.81 0.11 (4) &0.47 0.07 (3) &0.27          &    \\
Ni\,{\sc i} &-0.03 0.06 (17)&0.29 0.08 (15)&0.26 0.07 (20)&0.20 0.09
(19)&0.31 0.16 (13)&0.55 0.10 (11)&0.36 0.08 (18)&0.00 0.09 (13)&    \\
\noalign{\smallskip}
\hline\noalign{\smallskip}
\label{abundances}
\end{tabular}}
\end{table*}

\begin{table}
\caption[]{Comparison of [Fe/H] from Taylor (1996) and this
work. We
also give, in column 3 and 4, the VSL and SMR status for the
stars according to Taylor (1996).
} 
\begin{tabular}{llllllrrrrrrrrrrc}
\hline\noalign{\smallskip}
 ID & VSL & SMR  &[Fe/H] &  [Fe/H]  \\ 
  &     &      & (Taylor) & This work \\
\noalign{\smallskip}
\hline\noalign{\smallskip}
 HD 10780  & no   &      &  0.396  &-0.02 \\
 HD 32147  & yes  &      &$>0.1$   &0.28 \\
 HD 99491  & marg &      & 0.115   &0.20 \\
 HD 104304 & marg &      &0.326    &0.16\\
 HD 121370 &      & 95\% & 0.305   &0.25\\
 HD 145675 & marg & 98\% & 0.38    &0.47 \\
 HD 182572 &      & 98\% &$>$0.341 &0.35\\
 HD 196755 &      &      &0.500    &0.02 \\
\noalign{\smallskip}
\hline\noalign{\smallskip}
\label{iron_abundnaces}
\end{tabular}
\end{table}

The stellar abundances derived in this study are summarized in Table
\ref{abundances}.
We will discuss the abundance determination for each element separately. 
For some elements only one or a couple of lines have been used and the 
results are therefore more tentative than firm. The number of lines used 
for each element are also indicated in the table. 

\paragraph{Iron abundances}

Iron abundances are derived from a large number of lines, 28 to 44
lines per star, which means that the errors in the mean are very small,
typically less 0.02 dex. Thus, the error in Fe abundances is negelible 
in the error budget for the abundance ratios. 

In Table \ref{iron_abundnaces} we compare the iron abundances in this
study  and those quoted by Taylor (1996). For HD 32147, HD 99491, HD
121370, HD 145675, and  HD 182572 their SMR status is confirmed. HD
104304 is a marginal case and HD 10780 and HD 196755 are shown to not
be SMR stars.

\paragraph{Oxygen} 
The forbidden O\,{\sc i} line at 6300 {\AA} was only measured in two
stars, HD  99491 and HD 104304; they have [O/Fe] of 0.05 and -0.17
dex, respectively.  The errors in the O abundances are dominated by
the measurement error for the [O\,{\sc i}] line. Our spectra have S/N of
$\sim 250$, which translates into an  uncertainty of $\sim$0.15 dex in
[O/Fe]. Thus, our observed [O/Fe] are well within this scatter, and
our data follow the trends found in  Nissen \& Edvardsson (1992) and
Feltzing \& Gustafsson (1998).

Three of our stars have useful observations of the triplet lines around
777 nm. For those three stars we get [O/Fe] = 0.01, 0.28, and 0.09 dex
respectively. Line-to-line scatter is 0.1 dex or less for these stars
which means that formal
errors are less than 0.1 dex for all three stars. These oxygen abundances
should be fairly reliable as we are dealing with stars that are similar 
to the Sun and our study is differential. Edvardsson et al. (1993) found 
a good correlation between oxygen abundances derived from the 
forbidden
line and those derived from the triplet. Note, however, that Feltzing \&
Gustafsson (1998) found no such correlation for their very metal-rich
sample. 
Thus, in conclusion, the [O/Fe] for HD 196755 derived from the triplet 
should be robust while the [O/Fe] for HD 145675 and HD 182572 are more 
uncertain in term of possible NLTE effects.

\paragraph{Sodium} 
Only one Na\,{\sc i} line was available for analysis in our spectra, but
it 
has been widely used in other abundance studies.  Therefore, we are 
confident that it is giving us reliable Na abundances.

\paragraph{Aluminum} 
Aluminum shows a somewhat puzzling behaviour. Both
Edvardsson et al. (1993) and Feltzing \& Gustafsson (1998) found [Al/Fe] 
to be solar for all stars with [Fe/H] $>0.0$ dex. However, three of our
stars 
show unexpectedly high [Al/Fe] abundances. The line-to-line scatter is
small 
for all stars with all three lines measured. 

\paragraph{Silicon} 
A flat distribution with some internal scatter is found.
The line-to-line scatter is, as in Feltzing \& Gustafsson (1998), on the 
larger side for the number of lines used.

\paragraph{Calcium} For three stars we observe several lines of Ca\,{\sc i}.
These 
stars all show [Ca/Fe] close to the solar value, as expected. For the
other stars
only one or two lines were available and the results are therefore
uncertain.

\paragraph{Sulfur}

Our linelist contains 2 S\,{\sc i} lines, however, for those stars where we
could determine
S abundances only one line was available in each star. Our abundances are
therefore
uncertain. We note that [S/Fe] appears somewhat high. 

\paragraph{Scandium}

Sc abundances were derived from both Sc\,{\sc i} and  Sc\,{\sc ii} lines.
The 
[Sc/Fe] values fall within the range expected from Feltzing \& Gustafsson 
(1998) and the trend in our [Sc/Fe] data is flat at around 0.1-0.15 dex.
We 
note though that our most metal rich stars are overabundant in Sc in 
contrast with Feltzing \& Gustafsson (1998) which show a tendency for the 
most metal-rich stars to be underabundant.

\paragraph{Titanium} 
An overall flat trend is found for our stars in accordance with previous 
studies, i.e. Feltzing \& Gustafsson (1998). HD 32147 stands out with an 
extremely high Ti abundance, perhaps indicative of an underestimated
temperature 
(see Thor\'en \& Feltzing 2000).

\paragraph{Chromium} 
The behaviour is the same as found by Feltzing \& Gustafsson (1998),
except 
for HD 32147. Essentially, [Cr/Fe] has a flat trend above solar 
metallicity.

\paragraph{Cobalt} 
We find a large scatter in the [Co/Fe] abundances. 
The origin of this large scatter is not entirely clear, but it is 
accompanied by a large line-to-line scatter as well, and thus the 
reason may be sought among the selection of stellar lines. 

\paragraph{Nickel}
A large number of Ni\,{\sc i} lines were used, both to determine the
effective 
temperature as well as the Ni abundances. All our stars have virtually
solar [Ni/Fe] in agreement with  Edvardsson et al. (1993) and Feltzing \&
Gustafsson (1998) for stars in this [Fe/H] range. Interestingly, we find 
that our most metal-rich star, HD 145675, shows a slightly enhanced 
[Ni/Fe]. Tentative enhancements of Ni in the most metal-rich stars can
be found also in the studies by Feltzing \& Gustafsson (1998) and 
Thor\'en \& Feltzing (2000). All three studies adopted  different 
techniques for derivation of stellar parameters, and thus the result 
seems to be significant. However, further investigations should be 
undertaken. 

\section{Ages of SMR stars and the Age-metallicity relation}

\begin{figure}
\resizebox{\hsize}{!}{\includegraphics{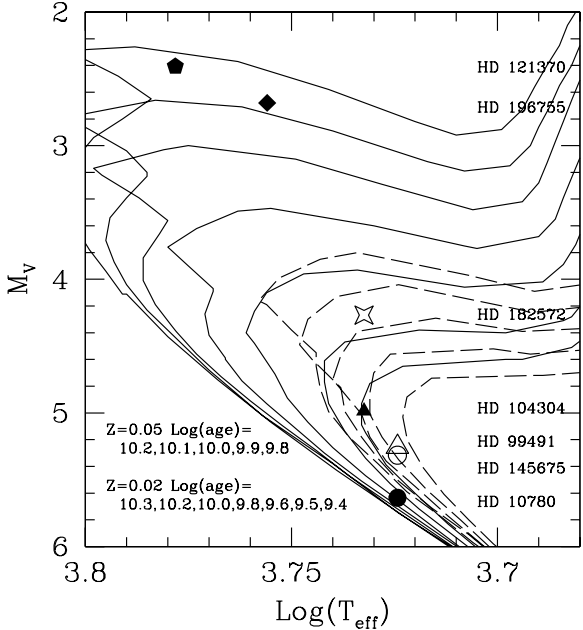}}
\caption[]{Diagram used to estimate ages for our program stars. Isochrones
are from Bertelli et al. (1994). Full curves correspond to Z=0.02, and the
dashed  curves to Z=0.05 isochrones, respectively. Filled symbols denote stars
with ${\rm [Fe/H]}\leq 0.20$ dex and open circles stars with  ${\rm [Fe/H]}>
0.20$ dex. HD 32147 is too cool to show on the diagram. 
}
\label{iso.fig}
\end{figure}

\begin{figure}
\resizebox{\hsize}{!}{\includegraphics[angle=-90]{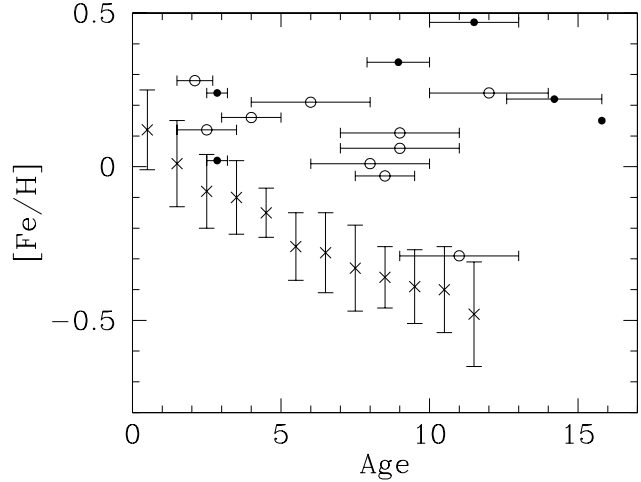}}
\caption[]{A comparison of our isochrone ages with the general
age-metallicity
relation derived by Rocha-Pinto et al. (2000). $\bullet$ refer to stars 
in this study, $\bigcirc$ to stars from Gonzalez, see Table
\ref{velage.tab}.
$\times$ refers to the age-metallicity relation from Rocha-Pinto et al. 
where the error bars give the 1 $\sigma$ 
scatter around the mean [Fe/H] for each age bin. 
}
\label{ages2.fig}
\end{figure}

An age-metallicity relation among dwarf stars in the solar
neighbourhood is a  key observable that models of galactic chemical
evolution must match. The most  important recent studies  include
Edvardsson et al. (1993), Carraro et  al. (1998), and Rocha-Pinto et
al. (2000). The first two studies use the same  [Fe/H], as derived in
Edvardsson et al. (1993) from detailed abundance analysis. Carraro et
al. (1998) make use of the age determinations  done for Edvardsson et
al. (1993) sample post Hipparcos (Ng \& Bertelli 1998).
Essentially, their data show a declining trend such that  more
metal-poor stars are older. However, the intrinsic scatter appears
large in both age and [Fe/H] and  a unique age-metallicity relation
may not be present. The study by Rocha-Pinto et al. (2000) used a
different technique to determine ages,  chromospheric activity. They
arrive at the conclusion that there exists  a unique age-metallicity
relation in the solar  neighbourhood. The scatter  in both age and
metallicity are found to be small for all ages and metallicities (see 
their Fig. 13).

SMR stars are rare and therefore none of the studies discussed contain
large numbers of them, in fact e.g. the  Edvardsson et al. (1993)
sample was selected with an upper limit in  metallicity near 0.2
dex. Such a bias is not present in the Rocha-Pinto et  al. (2000)
sample, and they have a few stars of up to $\sim 0.3$ dex (their Fig.
13). It is therefore valuable to derive ages for our small sample of
stars and compare them to that of the general age-metallicity
relations found in previous studies.

We have simply estimated the ages of the stars by plotting them in the
M$_V - T_{eff}$ plane and using the Bertelli et al.  (1994)
isochrones, Fig. \ref{iso.fig}. The ages were estimated by eye. The
correct isochrones were chosen depending on the [Fe/H] for each star
as derived in this study.   In order to see if the age-metallicity relation
appears unique also for the most metal-rich stars, we compare our data
and the ages from the several papers by Gonzalez and co-workers, see
 Table
\ref{velage.tab},
with the age-metallicity relation found in Rocha-Pinto et al. (2000) in
Figure \ref{ages2.fig}. 

A possible error source in the age determination of SMR stars is the
presence of planets. Gonzalez
(1998) noted that if one or several  planets have been engulfed by a
star, then its [Fe/H] may increase by up to around 0.10 dex. If this
has happened, then the abundances and age for  a polluted star will no
longer represent it's true age and abundances. However, such a change in
metallicity would still not turn a $\sim 10$ Gyr star into a star of
only a few Gyr, as required to fit into a general age-metallicity relation.

We note that our sample is not complete or in any other
way well-defined. However, it proves also  that there exist stars that are
both very old and at the same time very metal-rich, also
taking the errors in the ages into account. This casts doubts
on the possibility of defining a one-to-one relation between age and
metallicity among the  solar neighbourhood stars.

\section{Velocity data}

\begin{table*}
\caption[]{Velocity data and age determinations. The top half of the
table contains the stars in this study and the lower section stars
from  Gonzalez (1999) and Gonzalez \& Laws (2000). Age estimates,
using isochrones,  from this study are given in column 6 top half. The
age estimates for the  stars in the second section are also based on
the Bertelli et al. (1994)  stellar isochrones, with typical
uncertainties of 1 to 2 Gyrs. In column 7  we reproduce the Ca {\sc
ii} activity index ages from Gonzalez (1998).  In the last column we
give [Fe/H] as determined in this study or from  Gonzalez (1999) and
Gonzalez \& Laws (2000). }
\begin{tabular}{lrrrrrrrrrrrrr}
\hline\noalign{\smallskip}
ID  &Name & $U_{\rm LSR}$ & $V_{\rm LSR}$ & $W_{\rm LSR}$
& Age (isochrones) & Age (Ca {\sc ii}) & [Fe/H]\\ 
      &  & (km/s) & (km/s) & (km/s) & (Gyr) & (Gyr) \\
\noalign{\smallskip}
\hline\noalign{\smallskip}
  HD10780 &      & -14.8& -11.3&  1.6& --	      &  &-0.02 &	\\ 
  HD32147 &      &  10.7& -51.8& -6.3& --	      &  &0.28  &	\\ 
  HD99491 &      & -49.8& -4.1 & -8.7& 12.6 -- 15.8&  &0.22  &	\\ 
 HD104304 &      &  31.6& -10.2& -9.2& $\leq$15.8  &  &0.15  &	\\ 
 HD121370 &      &  19.2& -12.0&  4.6& 2.5 -- 3.2  &  &0.24  &	\\ 
 HD145675 &14 Her&  35.6&  -2.0& -2.9& 10 -- 13.0  &  &0.47  &	\\ 
 HD182572 &      &-106.7& -25.2&-13.5& 7.9 -- 10.0 &  &0.34  &	\\ 
 HD196755 &      &-48.1 &  29. &-11.7& 2.5 -- 3.2  &  &0.02  &	\\ 
\noalign{\smallskip}
\hline\noalign{\smallskip}
  HD9826  &$\upsilon$ And &  4.3  & -34.1 &   0.6 &  2.7      &  & 0.12 &
\\
  HD75732 &$\rho^{1}$ Cnc & -27.3 & -13.2 &  -0.9 &           &5 & 0.45
&\\
  HD75289 &          &  31.1 & -12.4 & -14.5 &  2.1      &  & 0.28 & \\
  HD95128 &47 UMa    & -14.7 &   2.6 &  8.8  &  6.3      &7 & 0.01 &\\
 HD117176 &70 Vir    &  23.2 & -46.9 &  3.4  &  8        &9 & -0.03& \\
 HD120136 &$\tau$ Boo& -23.5 & -13.8 &  0.3  &  1        &  & 0.32 & \\
 HD143761 &$\rho$ CrB&  64.1 & -30.7 & 28.5  &  12.3     &  & -0.29& \\
 HD186408 &16 Cyg A  &  27.6 & -23.6 &  7.2  &  9.0      &  & 0.11 & \\
 HD186427 &16 Cyg B  &  27.1 & -24.7 &  5.4  &  9.0      &7 & 0.06 & \\
 HD187123 &          &  11.6 & -10.6 & -36.4 &  5.5      &  & 0.16 &\\
 HD217014 &51 Peg    &  -5.6 & -24.2 & 22.3  &  6.0      &10& 0.21 & \\
 HD210277 &          &  12.4 & -46.8 &  3.0  &  8.5      &  & 0.24 & \\
\noalign{\smallskip}
\hline\noalign{\smallskip}
\label{velage.tab}
\end{tabular}
\end{table*}

Spatial velocity data were calculated using the Hipparcos parallaxes
and  proper motions. Radial velocities were taken from
Barbier-Brossat et al.  (1994). For our stars the
uncertainties in the parallaxes are  small, less than 3\% of the
parallax, Table \ref{parameters}. Data were  also obtained for the
stars from the Gonzalez (1999) compilation. The  velocities are
presented in Table \ref{velage.tab}. Note that we here quote the
velocities relative to the local standard of rest (LSR) and  Gonzalez
(1999) quoted velocities relative to the sun.

From Table \ref{velage.tab} we see that all the stars have
$W$-velocities well below the $\sigma_W$ of the general population of
stars with similar $B-V$. Fig. 5 in Dehnen \&
Binney (1998) illustrates how $\sigma_{U,V,W}$ varies with $B-V$.
 Also, most of the stars in Table
\ref{velage.tab} have both $V$ and $U$-velocities well below 1
$\sigma$ for the general population.  We have
quantified this by calculating the probabilities that any one of our
stars belongs to either the thin or the thick disk by using a model
where 94\% of the solar neighbourhood  stars belong to a thin disk
with $\sigma_U=35$, $\sigma_V=25$, and $\sigma_W=17$ km/s and the
remaining 6\% to the thick disk with $\sigma_U=70$, $\sigma_V=50$,
and $\sigma_W=34$ km/s. Only one of our stars, HD 182572, has a
probability that it belongs to the thick disk larger than  that
it should belong to the thin. We estimate that, given the galactic
model, this star has 75\% chance of belonging to the thick disk.
Thus, we conclude that  our SMR and
planet-bearing stars samples the thin disk.

Fuhrmann (1998) found that
stars with thick  disk kinematics were enhanced in [Mg/Fe] as compared
to thin disk stars at the same metallicity. We
have not measured Mg lines in our spectra. We did, however, measure Si, and our 
abundance result for HD 182572 gives [Si/Fe] $= 0.16 \pm 0.09$. 
Compared with the general trend of [Si/Fe] for metal-rich
stars in Feltzing \& Gustafsson (1998), this is above the mean;
however, their data exhibit a large scatter. 
We have also determined Ca abundances for this 
star, however, only one line was available. This line seems to give 
fairly low Ca abundances in all of the stars with more than
two lines observed and may thus be underestimating the true Ca abundance 
in this star.  Note that it is not inconsistent that we also
find thin disk stars with the same Si abundance as, if HD 182572 is a
thick disk star, then it might be showing us the abundance trend after
the decline in [X/Fe], where X is either O or an $\alpha$-element,
 sets in due to  increasing relative contribution
of SNIa.

\section{SMR - planet connection}

Several of the nearby SMR star candidates listed by Taylor (1996;
Table 4)  have been found to harbour planets. In particular, the
planet-bearing stars HD 75732, HD 145675, and HD 217014 are included
in  Taylor's list of 29 SMR class IV-V star candidates. Not all the
stars in  his list have been searched for planets as of yet, so the 
fraction of SMR star  candidates with planets may increase. Butler et
al. (2000) provide independent  confirming evidence for a
preponderance of planets among metal-rich stars;  they note that of
their 600 Keck targets, half of which are metal-poor,  2 planets have
been found around metal-poor stars and 10 around metal-rich
stars. Also, several planet-bearing stars not in Taylor's list have
recently  been found to be likely SMR stars. Examples in this group
include HD 120136,  HD 217107, and HD 210277 (Gonzalez 2000). It
appears that the Doppler planet  search method is also an efficient
detector of new SMR star candidates!

Gonzalez et al. (1999) suggested that BD $-10{\degr} 3166$ be searched for
planets on  the basis of its similarity to HD 75732 and HD 145675
(similar $T_{\rm eff}$  and [Fe/H]). Butler et al. (2000) reported on
the detection of a planet  around BD $-10{\degr} 3166$, which they had placed on
their monitoring program as a  result of Gonzalez et al.'s suggestion.
Another star, HD 89744, was suggested  to Geoff Marcy as a
planet-bearing candidate by one of us (G.G.) on the  basis of its high
[Fe/H] and low [C/Fe]. Sylvain et al. (2000) announced the  discovery
of a planet around this star (note: they began observing this star
about 2 years before Gonzalez's prediction).  The successful
prediction of  the presence of planets around two stars provides
strong independent  confirming evidence of the planet - SMR star
connection.  The low [C/Fe]  values seen among stars with planets is
the first convincing evidence of a  trend with abundance ratios. The
physical significance of this trend is not yet known.

\section{Conclusions}

We have presented detailed abundance analyses at high resolution for 8
possible SMR dwarf stars and subgiants. Four of these stars have
previously  been studied at high resolution; our results in general
agree well with them.  For the remaining four stars this is, to the
best of our knowledge, the first  study of this sort.

We find in particular that:

\begin{itemize}

\item HD32147, HD99491,  HD121370, HD145675, HD182572 
all have ${\rm [Fe/H]} \geq 0.2$ dex, the lower limit for 
super metal rich status as defined by Taylor (1996)

\item HD104304 presents a marginal case

\item HD10780 and HD196755 are found to have solar iron abundances and are
thus not SMR stars

\item some metal-rich, and in particular SMR, stars are old, showing 
that the large scatter in [Fe/H] at a given age among nearby solar type 
stars exists at all ages and that a one-to-one relation between age and 
metallicity among the solar neighbourhood stars may not exist for all
metallicities.

\item that metal-rich stars are mainly confined to the galactic plane,
however, one star in our sample appears to be a thick disk candidate of 
extremely high metallicity

\item there exists a correlation between SMR-ness and the presence of
planets
\end{itemize}

Further investigations should be undertaken to prove the possibility
that [Ni/Fe] is starting to increase at the highest metallicities as 
well as the very real possibility that there exists extreme thick disk
stars with very high metallicities. In particular, $\alpha$-element 
abundances should be carefully studied for such candidates.

\begin{acknowledgements}
SF acknowledges financial support from the Swedish Natural Research
Council
under their postdoc program. GG acknowledges financial support from the 
Kennilworth Fund of the New York Community Trust.

Johan Holmberg at Lund Observatory is thanked for providing stellar
velocities.
\end{acknowledgements}

\begin{appendix}

\begin{table}
\caption[]{Line data. }
\begin{tabular}{lrrrrrrl}
\hline\noalign{\smallskip}
{ $\lambda$} & { $\chi_l$} & {$ \log gf$} & {$\delta \Gamma_6$} & 
{  $\Gamma _{\rm rad}$} &  Note  \\ 
{[\AA] }      &[eV]       &            &                     &[s$^{-1}$]
&     \\ 
\noalign{\smallskip}
\hline\noalign{\smallskip}
\multicolumn{6}{l} {{\bf C\,{\sc i}}; $\log \epsilon_\odot = 8.56$}&\\ 
  6587.61   &   8.53  &   -1.246  &   2.50 & 1.00e+08 & \\ 
\multicolumn{6}{l} {{\bf O\,{\sc i}}; $\log \epsilon_\odot = 8.93$}&\\ 
  6300.310 &   0.00   &   -9.75   &    2.50  & 1.00e+08  & \\  
  7771.95  &   9.14   &    0.26   &    2.50  & 1.00e+08  &  \\
  7774.17  &   9.14   &    0.12   &    2.50  & 1.00e+08  & \\ 
  7775.35  &   9.14   &   -0.11   &    2.50  & 1.00e+08  & \\ 
\multicolumn{6}{l} {{\bf Na\,{\sc i}}; $\log \epsilon_\odot = 6.33$}&\\ 
  6154.23    &   2.10   &   -1.58    &   2.10 &  7.08e+07    & \\ 
\multicolumn{6}{l} {{\bf Al\,{\sc i}}; $\log \epsilon_\odot = 6.47$}&\\ 
  6698.67    &   3.143  &   -1.89  &     2.50 &  3.02e+08   & \\ 
  7835.30    &   4.022  &   -0.78   &    2.50 &  7.94e+07   & \\ 
  7836.13    &   4.022   &  -0.60  &     2.50 &  7.94e+07   & \\ 
\multicolumn{6}{l} {{\bf Si\,{\sc i}}; $\log \epsilon_\odot =7.55  $}&\\ 
  5622.22   &  4.930   &  -2.95    &   2.50  & 1.95e+08  & \\ 
  5665.55   &  4.920   &  -2.02    &   2.50  & 1.95e+08  & \\
  5793.07   &  4.930   &  -1.95    &   2.50  & 1.95e+08  & \\ 
  6125.02   &  5.614   &  -1.55    &   2.50  & 1.00e+08  &  \\ 
  6142.48   &  5.619   &  -1.50    &   2.50  & 1.00e+08  &  \\ 
  6155.69   &  5.619   &  -2.43    &   2.50  & 1.00e+08  &  \\  
  6237.31   &  5.614   &  -1.15    &   2.50  & 1.00e+08  & \\ 
  6721.84   &  5.863   &  -1.16    &   2.50  & 1.00e+08  & \\ 
  6741.62   &  5.984   &  -1.63    &   2.50  & 2.69e+07  & \\ 
  6848.58   &  5.863   &  -1.65    &   2.50  & 1.00e+08  & \\ 
  7455.37   &  5.964   &  -2.00    &   2.50  & 1.00e+08  & \\ 
  7760.62   &  6.206   &  -1.36    &   2.50  & 1.00e+08  & \\ 
  7799.99   &  6.181   &  -0.77    &   2.50  & 1.00e+08  & \\ 
\multicolumn{6}{l} {{\bf S\,{\sc i}}; $\log \epsilon_\odot = 7.21$}&\\ 
  6743.531  &  7.866  & -0.84   &  2.50 &3.80e+07   & \\ 
  6757.171  &  7.870  & -0.53   &  2.50 &3.89e+07   & \\ 
\multicolumn{6}{l} {{\bf Ca\,{\sc i}}; $\log \epsilon_\odot =6.36 $}&\\ 
  5512.980  &  2.933   &  -0.66  &     2.50  & 2.80e+08  & \\ 
  5581.965  &  2.523   &  -0.87  &     2.50  & 7.13e+07  & \\ 
  5867.562  &  2.933   &  -1.61  &     2.50  & 2.62e+08  & \\ 
  6166.439  &  2.521   &  -1.22  &     2.50  & 1.86e+07  & \\ 
  6169.042  &  2.523   &  -0.90  &     2.50  & 1.86e+07  & \\
  6169.563  &  2.526   &  -0.67  &     2.50  & 1.88e+07  & \\ 
  6455.598  &  2.523   &  -1.48  &     2.50  & 4.64e+07  & \\ 
  6464.673  &  2.526   &  -2.36  &     2.50  & 4.64e+07  & \\ 
  6471.662  &  2.526   &  -0.98  &     2.50  & 4.42e+07  & \\ 
\multicolumn{6}{l} {{\bf Sc\,{\sc i}}; $\log \epsilon_\odot =3.10 $}&\\ 
  5484.626   & 1.851 &   0.37   &  1.50 &1.46e+08    & \\
\multicolumn{6}{l} {{\bf Sc\,{\sc ii}}; $\log \epsilon_\odot =3.10 $}&\\ 
  5526.790  &  1.768  &  0.09  &   1.50& 2.16e+08   & \\
  6300.698  &  1.507  & -2.01  &   1.50& 2.31e+08   & \\
  6604.601  &  1.357  & -1.16  &   1.50& 1.46e+08   & \\
\multicolumn{6}{l} {{\bf Ti\,{\sc i}}; $\log \epsilon_\odot = 4.99$}&\\ 
  5490.148  &  1.460  &   -0.98  &     2.50  & 1.43e+08     & \\
  5739.469  &  2.249  &   -0.78  &     2.50  & 6.60e+07     & \\
  5739.978  &  2.236  &   -0.74  &     2.50  & 6.53e+07     & \\
  5823.686  &  2.267  &   -1.05  &     2.50  & 6.53e+07     & \\
  5832.473  &  3.337  &   -0.28  &     2.50  & 1.45e+08     & \\
  5866.451  &  1.067  &   -0.83  &     2.50  & 4.40e+08     & \\
  6091.171  &  2.267  &   -0.43  &     2.50  & 8.50e+07     & \\
  6092.792  &  1.887  &   -1.45  &     2.50  & 1.27e+08     & \\
  6098.658  &  3.062  &   -0.07  &     2.50  & 5.43e+07     & \\
  6126.216  &  1.067  &   -1.41  &     2.50  & 9.93e+06     & \\
  6599.105  &  0.900  &   -2.07  &     2.50  & 1.22e+06     & \\
  6743.122  &  0.900  &   -1.73  &     2.50  & 6.93e+05     & \\
  6745.547  &  2.236  &   -1.36  &     2.50  & 1.44e+08     & \\
\noalign{\smallskip}\hline
\end{tabular}
\end{table}

\begin{table}
\begin{tabular}{lrrrrrrl}
\hline\noalign{\smallskip}
{ $\lambda$} & { $\chi_l$} & {$ \log gf$} & {$\delta \Gamma_6$} & 
{  $\Gamma _{\rm rad}$} &  Note  \\ 
{[\AA] }      &[eV]       &            &                     &[s$^{-1}$]
&     \\ 
\noalign{\smallskip}
\hline\noalign{\smallskip}
\multicolumn{6}{l} {{\bf V\,{\sc i}}; $\log \epsilon_\odot = 4.00 $}&\\ 
  5670.853   &   1.081   &  -0.46    &   2.50  & 5.23e+06    & \\
  5727.048   &   1.081   &  -1.22    &   2.50  & 7.57e+08    & \\
  5727.652   &   1.051   &  -0.90    &   2.50  & 6.15e+07    & \\
  5830.675   &   3.113   &   0.61    &   2.50  & 1.83e+08    & \\
  6039.722   &   1.064   &  -0.72    &   2.50  & 3.98e+07    & \\
  6090.214   &   1.081   &  -0.15    &   2.50  & 3.98e+07    & \\
  6111.645   &   1.043   &  -0.795   &   2.50  & 3.90e+07    & \\
  6135.361   &   1.051   &  -0.766   &   2.50  & 3.90e+07    & \\  
  6150.157   &   0.301   &  -1.54    &   2.50  & 7.81e+05    & \\
  6199.197   &   0.287   &  -1.48    &   2.50  & 3.25e+06    & \\
  6224.529   &   0.287   &  -1.84    &   2.50  & 1.22e+06    & \\
  6251.827   &   0.287   &  -1.45    &   2.50  & 3.07e+07    & \\
  6256.887   &   0.275   &  -2.17    &   2.50  & 2.94e+06     & \\ 
  6452.341   & 1.195  & -0.836  &  2.50& 3.99e+07    & \\
\multicolumn{6}{l} {{\bf Cr\,{\sc i}}; $\log \epsilon_\odot = 5.67$}&\\ 
  5628.621   & 3.422   &  -0.83   &    2.50 &  6.52e+07      & \\
  5664.555   & 3.826   &  -0.87   &    2.50 &  4.80e+07      & \\
  5783.886   & 3.322   &  -0.26   &    2.50 &  9.98e+07      & \\
  5787.965   & 3.322   &  -0.21   &    2.50 &  1.00e+08      & \\
  6630.005   & 1.030   &  -3.46   &    2.50 &  2.40e+07      & \\   
  6669.255   & 4.175   &  -0.47   &    2.50 &  3.66e+07      & \\
  7400.226   & 2.900   &  -0.171  &    2.50 &  6.76e+07      & \\
\multicolumn{6}{l} {{\bf Cr\,{\sc ii}}; $\log \epsilon_\odot =5.67 $}&\\ 
  5502.067   & 4.168  & -1.99   &  2.50 &2.55e+07    & \\ 
\multicolumn{6}{l} {{\bf Fe\,{\sc i}}; $\log \epsilon_\odot = 7.51$}&\\ 
  5491.832  &  4.186   &  -2.14    &   2.00 &  1.44e+08  & \\ 
  5494.463  &  4.076   &  -1.85    &   2.00 &  2.90e+07  & \\ 
  5522.447  &  4.209   &  -1.43    &   2.00 &  8.97e+07  & \\ 
  5539.280  &  3.642   &  -2.49    &   2.00 &  2.60e+07  & \\ 
  5543.936  &  4.217   &  -1.10    &   2.00 &  2.39e+08  & \\ 
  5560.212  &  4.434   &  -1.09    &   2.00 &  1.64e+08  & \\ 
  5577.030  &  5.033   &  -1.45    &   2.00 &  6.89e+08  & \\ 
  5579.340  &  4.231   &  -2.29    &   2.00 &  2.55e+08  & \\ 
  5607.664  &  4.154   &  -2.18    &   2.00 &  3.50e+08  & \\ 
  5608.972  &  4.209   &  -2.31    &   2.00 &  8.89e+07  & \\ 
  5611.360  &  3.635   &  -2.91    &   2.00 &  1.25e+08  & \\ 
  5618.633  &  4.209   &  -1.34    &   2.00 &  1.05e+08  & \\ 
  5619.595  &  4.386   &  -1.48    &   2.00 &  1.78e+08  & \\ 
  5636.696  &  3.640   &  -2.56    &   2.00 &  3.90e+07  & \\ 
  5638.262  &  4.220   &  -0.88    &   2.00 &  1.94e+08  & \\ 
  5646.684  &  4.260   &  -2.40    &   2.00 &  1.00e+08  & \\ 
  5651.469  &  4.473   &  -1.76    &   2.00 &  1.62e+08  & \\ 
  5652.010  &  4.218   &  -1.82    &   2.00 &  2.36e+08  & \\ 
  5741.848  &  4.256   &  -1.66    &   2.00 &  2.11e+08  & \\ 
  5793.689  &  4.593   &  -1.29    &   2.00 &  5.38e+07  & \\ 
  5811.914  &  4.143   &  -2.38    &   2.00 &  3.76e+07  & \\ 
  5814.807  &  4.283   &  -1.84    &   2.00 &  2.11e+08  & \\ 
  5852.219  &  4.548   &  -1.21    &   2.00 &  1.90e+08  & \\ 
  5855.077  &  4.608   &  -1.54    &   2.00 &  1.91e+08  & \\ 
  6034.035  &  4.312   &  -2.30    &   2.00 &  1.59e+08  & \\ 
  6089.580  &  4.580   &  -1.30    &   2.00 &  9.27e+07  & \\ 
  6094.374  &  4.652   &  -1.58    &   2.00 &  1.92e+08  & \\ 
  6096.665  &  3.984   &  -1.81    &   2.00 &  4.53e+07  & \\ 
  6105.131  &  4.548   &  -1.91    &   2.00 &  1.00e+08  & \\ 
  6151.618  &  2.176   &  -3.329   &   2.00 &  1.55e+08  & \\ 
  6157.728  &  4.076   &  -1.28    &   2.00 &  5.02e+07  & \\ 
  6159.378  &  4.607   &  -1.85    &   2.00 &  1.92e+08  & \\ 
  6165.360  &  4.143   &  -1.524   &   2.00 &  8.77e+07  & \\ 
  6173.336  &  2.223   &  -2.910   &   2.00 &  1.67e+08  & \\ 
  6187.990  &  3.943   &  -1.65    &   2.00 &  4.60e+07  & \\ 
\noalign{\smallskip}\hline
\end{tabular}
\end{table}

\begin{table}
\begin{tabular}{lrrrrrrl}
\hline\noalign{\smallskip}
{ $\lambda$} & { $\chi_l$} & {$ \log gf$} & {$\delta \Gamma_6$} & 
{  $\Gamma _{\rm rad}$} &  Note  \\ 
{[\AA] }      &[eV]       &            &                     
&[s$^{-1}$]                &     \\ 
\noalign{\smallskip}
\hline\noalign{\smallskip}
  6200.313  &  2.608   &  -2.48    &   2.00 &  1.03e+08  & \\ 
  6226.736  &  3.883   &  -2.09    &   2.00 &  5.42e+07  & \\ 
  6229.228  &  2.845   &  -2.885   &   2.00 &  1.45e+08  & \\ 
  6380.743  &  4.186   &  -1.366   &   2.00 &  7.34e+07  & \\ 
  6385.718  &  4.733   &  -1.82    &   2.00 &  2.34e+08  & \\ 
  6419.950  &  4.733   &  -0.27    &   2.00 &  2.31e+08  & \\ 
  6591.313  &  4.593   &  -1.99    &   2.00 &  1.40e+08  & \\ 
  6608.026  &  2.279   &  -3.94    &   2.00 &  1.66e+08  & \\ 
  6653.853  &  4.154   &  -2.38    &   2.00 &  2.09e+08  & \\ 
  6703.567  &  2.758   &  -3.03    &   2.00 &  1.03e+08  & \\ 
  6710.319  &  1.485   &  -4.79    &   2.00 &  1.66e+07  & \\ 
  6726.661  &  4.607   &  -1.079   &   2.00 &  2.29e+08  & \\ 
  6733.151  &  4.638   &  -1.45    &   2.00 &  2.27e+08  & \\  
  6745.101  &  4.580   &  -2.17    &   2.00 &  4.78e+07  & \\
  6745.957  &  4.076   &  -2.74    &   2.00 &  3.79e+07  & \\ 
  6750.153  &  2.424   &  -2.641   &   2.00 &  7.69e+06  & \\
  6806.845  &  2.727   &  -3.11    &   2.00 &  1.02e+08  & \\
  6810.263  &  4.607   &  -1.026   &   2.00 &  2.30e+08  & \\
  6820.372  &  4.638   &  -1.16    &   2.00 &  2.22e+08  & \\
  7401.685  &  4.186   &  -1.609   &   2.00 &  7.03e+07  & \\
  7418.667  &  4.143   &  -1.476   &   2.00 &  1.06e+08  & \\
  7421.559  &  4.638   &  -1.73    &   2.00 &  2.50e+08  & \\
  7440.952  &  4.913   &  -0.662   &   2.00 &  4.97e+08  & \\
  7443.022  &  4.186   &  -1.66    &   2.00 &  3.48e+07  & \\
  7453.998  &  4.186   &  -2.36    &   2.00 &  1.45e+08  & \\ 
  7498.530  &  4.143   &  -2.13    &   2.00 &  1.16e+08  & \\
  7507.261  &  4.415   &  -0.972   &   2.00 &  9.77e+07  & \\
  7540.430  &  2.727   &  -3.77    &   2.00 &  9.31e+07  & \\
  7547.910  &  5.100   &  -1.247   &   2.00 &  6.38e+08  & \\
  7551.109  &  5.085   &  -1.57    &   2.00 &  6.34e+08  & \\
  7582.122  &  4.955   &  -1.64    &   2.00 &  1.00e+08  & \\
  7588.305  &  5.033   &  -1.07    &   2.00 &  6.35e+08  & \\
  7745.500  &  5.086   &  -1.149   &   2.00 &  6.37e+08  & \\ 
  7746.587  &  5.064   &  -1.276   &   2.00 &  6.31e+08  & \\
  7751.137  &  4.991   &  -0.775   &   2.00 &  6.44e+08  & \\
  7844.559  &  4.835   &  -1.70    &   2.00 &  2.34e+08  & \\
\multicolumn{6}{l} {{\bf Fe\,{\sc ii}}; $\log \epsilon_\odot = 7.51$}&\\ 
  5823.155  &  5.569   &  -2.88   &    2.00  & 3.00e+08   & not used  \\ 
  6149.258  &  3.889   &  -2.84   &    2.00  & 3.39e+09   &  \\ 
  6247.557  &  3.892   &  -2.39   &    2.00  & 3.38e+08   &  \\ 
  6369.462  &  2.891   &  -4.20   &    2.00  & 2.90e+08   &  \\ 
  6383.722  &  5.553   &  -2.10   &    2.00  & 4.09e+08   &  \\ 
  6416.919  &  3.892   &  -2.69   &    2.00  & 3.37e+08   & not used \\ 
  6432.680  &  2.891   &  -3.64   &    2.00  & 2.90e+07   &  \\ 
  6446.410  &  6.223   &  -1.88   &    2.00  & 5.22e+08   &  \\ 
  6456.383  &  3.903   &  -2.24   &    2.00  & 3.37e+08   &  \\ 
  7449.335  &  3.889   &  -3.168  &    2.00  & 4.09e+08   & not used \\ 
  7515.831  &  3.903   &  -3.462  &    2.00  & 4.09e+08   &  \\ 
\multicolumn{6}{l} {{\bf Co\,{\sc i}}; $\log \epsilon_\odot = 4.92$}&\\ 
  5647.234  &  2.280   &  -1.58   &    2.50  & 1.66e+08  &  \\
  6086.658  &  3.409   &   0.22   &    2.50  & 8.87e+07  &  \\
  6093.143  &  1.740   &  -2.37   &    2.50  & 2.08e+07  &  \\
  6188.996  &  1.710   &  -2.33   &    2.50  & 2.22e+07  &  \\
  6257.572  &  3.713   &  -1.08   &    2.50  & 6.89e+07  &  \\
  6429.906  &  2.137   &  -2.39   &    2.50  & 9.29e+05  &  \\
  6632.433  &  2.280   &  -1.87   &    2.50  & 6.46e+06  &  \\
  6814.942  &  1.956   &  -1.77   &    2.50  & 2.08e+07  &  \\
  7417.367  &  2.042   &  -1.97   &    2.50  & 2.22e+07  &  \\
\noalign{\smallskip}\hline
\end{tabular}
\end{table}

\begin{table}
\begin{tabular}{lrrrrrrl}
\hline\noalign{\smallskip}
{ $\lambda$} & { $\chi_l$} & {$ \log gf$} & {$\delta \Gamma_6$} & 
{  $\Gamma _{\rm rad}$} &  Note  \\ 
{[\AA] }      &[eV]       &            &                     
&[s$^{-1}$]                &     \\ 
\noalign{\smallskip}
\hline\noalign{\smallskip}
\multicolumn{6}{l} {{\bf Ni\,{\sc i}}; $\log \epsilon_\odot = 6.25$}&\\ 
  5494.876   &   4.105  &   -1.08  &     2.50 &  2.05e+08 &  \\
  5578.711   &   1.676  &   -2.68  &     2.50 &  5.43e+07 &  \\
  5628.335   &   4.089  &   -1.35  &     2.50 &  2.52e+08 &  \\
  5638.745   &   3.898  &   -1.72  &     2.50 &  1.27e+08 &  \\
  5643.072   &   4.165  &   -1.22  &     2.50 &  9.14e+07 &  \\
  6025.751   &   4.236  &   -1.69  &     2.50 &  1.92e+08 &  \\
  6039.296   &   4.236  &   -2.07  &     2.50 &  2.16e+08 &  \\
  6086.276   &   4.266  &   -0.51  &     2.50 &  2.54e+08 &  \\
  6108.107   &   1.676  &   -2.56  &     2.50 &  4.86e+07 &  \\
  6111.066   &   4.088  &   -0.86  &     2.50 &  1.46e+08 &  \\
  6128.963   &   1.676  &   -3.37  &     2.50 &  1.21e+07 &  \\
  6130.130   &   4.266  &   -0.963 &     2.50 &  2.78e+08 &  \\
  6133.963   &   4.088  &   -1.79  &     2.50 &  1.45e+08 &  \\
  6176.807   &   4.088  &   -0.27  &     2.50 &  1.45e+08 &  \\
  6177.236   &   1.826  &   -3.562 &     2.50 &  4.31e+07 &  \\
  6186.709   &   4.105  &   -0.922 &     2.50 &  2.06e+09 &  \\
  6204.600   &   4.088  &   -1.146 &     2.50 &  1.75e+08 &  \\
  6370.341   &   3.542  &   -1.957 &     2.50 &  1.48e+08 &  \\
  6378.247   &   4.154  &   -0.84  &     2.50 &  2.08e+08 &  \\
  6424.847   &   4.167  &   -1.354 &     2.50 &  2.13e+08 &  \\
  6586.308   &   1.951  &   -2.748 &     2.50 &  4.30e+07 &  \\
  6598.593   &   4.236  &   -0.94  &     2.50 &  1.92e+08 &  \\
  6643.629   &   1.676  &   -2.08  &     2.50 &  1.01e+08 &  \\
  6767.768   &   1.826  &   -2.24  &     2.50 &  1.03e+08 &  \\
  6772.313   &   3.658  &   -1.082 &     2.50 &  1.50e+08 &  \\
  7414.500   &   1.986  &   -2.22  &     2.50 &  1.03e+08 &  \\
  7522.758   &   3.658  &   -0.535 &     2.50 &  1.50e+08 &  \\
  7525.111   &   3.635  &   -0.64  &     2.50 &  1.19e+08 &  \\
  7574.043   &   3.833  &   -0.533 &     2.50 &  9.12e+07 &  \\
  7826.766   &   3.699  &   -1.856 &     2.50 &  5.98e+07 &  \\
\noalign{\smallskip}\hline
\end{tabular}
\label{linelist}
\end{table}

\end{appendix}

\end{document}